\documentclass[a4paper,11pt]{article}
\usepackage{pos}
\usepackage{siunitx}
\usepackage[nameinlink]{cleveref}
\usepackage{wrapfig}

\title{A novel approach for air shower profile reconstruction with dense radio antenna arrays using Information Field Theory}
\ShortTitle{Air Shower Profile Reconstruction with IFT}

\author*[a]{K.~Watanabe}
\author[b]{S.~Bouma}
\author[c]{J.D.~Bray}
\author[d,e]{S.~Buitink}
\author[d,e]{A.~Corstanje} 
\author[d]{V.~De Henau}
\author[d]{M.~Desmet}
\author[f]{E.~Dickinson}
\author[e]{L.~van Dongen}
\author[g,h,i]{T.A.~En{\ss}lin}
\author[j,k]{B.~Hare}
\author[l]{H.~He}
\author[e]{J.R.~H\"orandel}
\author[a,d]{T.~Huege}
\author[f]{C.W.~James}
\author[g,h]{M.~Jetti}
\author[b]{P.~Laub}
\author[a]{H.J.~Mathes}
\author[e,m]{K.~Mulrey}
\author[b,n]{A.~Nelles}
\author[a,o]{S.~Saha}
\author[j]{O.~Scholten}
\author[b]{S.~Sharma}
\author[c]{R.E.~Spencer}
\author[k]{C.~Sterpka}
\author[k]{S.~ter Veen}
\author[b]{K.~Terveer}
\author[p]{S.~Thoudam}
\author[q]{T.N.G.~Trinh}
\author[j,k]{P.~Turekova}
\author[a]{D.~Veberi\v{c}}
\author[r]{M.~Waterson}
\author[s,t]{C.~Zhang}
\author[u]{P.~Zhang}
\author[l,v]{Y.~Zhang}

\emailAdd{keito.watanabe@kit.edu}

\abstract{Reconstructing the longitudinal profile of extensive air showers, generated from the interaction of cosmic rays in the Earth's atmosphere, is crucial to understanding their mass composition, which in turn provides valuable insight on their possible sources of origin. Dense radio antenna arrays such as the LOw Frequency ARray (LOFAR) telescope as well as the upcoming Square Kilometre Array Observatory (SKAO) are ideal instruments to explore the potential of air shower profile reconstruction, as their high antenna density allows cosmic ray observations with unprecedented accuracy. However, current analysis approaches can only recover $X_\mathrm{max}$, the atmospheric depth at shower maximum, and heavily rely on computationally expensive simulations. As such, it is ever more crucial to develop new analysis approaches that can perform a full air shower profile reconstruction efficiently.

In this work, we develop a novel framework to reconstruct the longitudinal profile of air showers using measurements from radio detectors with Information Field Theory (IFT), a state-of-the-art reconstruction framework based on Bayesian inference. Through IFT, we are able to exploit all available information in the signal (amplitude, phase, and pulse shape) at each antenna position simultaneously and explicitly utilise models that are motivated through our current understanding of air shower physics. We verify our framework on simulated datasets prepared for LOFAR, showcasing that we can not only reconstruct the air shower profile with uncertainties in each atmospheric depth bin but also recover the reconstructed trace at each antenna position. Our framework demonstrates that radio measurements with dense antenna layouts such as LOFAR and SKAO have the capability to go beyond reconstruction of $X_\mathrm{max}$ and will thus aid in our understanding of the mass composition of cosmic rays. }

\ConferenceLogo{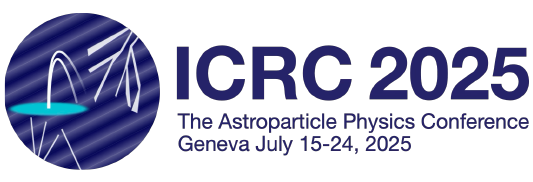}

\FullConference{39th International Cosmic Ray Conference (ICRC2025)\\
 15–24 July 2025\\
Geneva, Switzerland\\}
\begin{document}
\maketitle

\section{Introduction}

The origins of high-energy cosmic rays remain elusive, and understanding their primary mass composition is essential for uncovering the sources of these particles. Radio emission produced by cosmic ray air showers has proven to be an effective method for cosmic ray detection. Notably, the upcoming Square Kilometre Array Observatory (SKAO), with its SKA-Low array having over 60,000 antennas distributed across a \SI{1}{\kilo\metre\squared} core area, will greatly enhance our capacity to study cosmic rays \cite{Huege:2016jvc}. Current reconstruction techniques for mass composition in dense antenna arrays such as the LOw Frequency ARray (LOFAR) and SKA-Low rely on energy fluence measurements from numerous \texttt{CoREAS} simulations~\cite{Huege:2013vt} to estimate $X_\mathrm{max}$, the atmospheric depth at shower maximum~\cite{Buitink:2014, Corstanje:2025wbc}. However, additional insights into mass composition and the hadronic interaction models underlying the interpretation of measured data can be obtained from other parameters, including the width and asymmetry of the longitudinal profile of the air shower~\cite{Andringa:2011zz}. With its unprecedented precision in measuring cosmic ray air showers, SKA-Low has the capability to reconstruct more than just $X_\mathrm{max}$~\cite{Buitink:2023reh,Corstanje:2023uyg}, making it imperative to develop analyses that can extract this additional information from the radio emission.\par 

In this work, we showcase the development of a novel approach to reconstruct the longitudinal profile of air showers using Information Field Theory (IFT)~\cite{Ensslin:2018pno}. This method allows us to not only reconstruct shower parameters more than just $X_\mathrm{max}$, but also do so without explicitly relying on simulations. In our model, we parameterise the longitudinal profile as a Gaisser-Hillas function and utilise a fast forward model based on template synthesis~\cite{Desmet:2025ufy} to simulate the radio emission from air showers. We include the antenna response for antennas used in SKA-Low for a realistic detector description. With IFT, we infer the shower parameters from realistically simulated data from \texttt{CoREAS} simulations and demonstrate the reconstruction efficiency of the longitudinal profile of air showers on the basis of an idealised star-shaped antenna layout. Through our method, we highlight the potential of air shower profile reconstruction for dense radio detectors such as SKA-Low. 

\section{Model}

The model is constructed through IFT, a Bayesian reconstruction framework that can be applied to signal reconstruction problems~\cite{Ensslin:2018pno}. IFT allows us to infer parameters of a model from noisy data by combining domain and prior knowledge and likelihood functions motivated by our current physical understanding of air showers. Furthermore, its inference procedure based on variational inference makes it possible to model parameters defined at each atmospheric depth bin, making it an ideal tool for reconstructing the entire profile. The Bayesian nature of IFT also allows us to reconstruct the signal through the inferred shower parameters. IFT has already shown its applicability for reconstructing the electric field signal from air showers~\cite{Welling:2021cgl,Strahnz:2024tcw}.  \par 

To utilise the IFT framework, the forward model (prior + likelihood function) needs to be fully differentiable. This is automatically encoded in \texttt{JAX}, the language framework that is used by the numerical framework for IFT (\texttt{NIFTy},~\citep{niftyre}). Another necessary requirement is for the forward model to be computationally performant, as hundreds of instances of the forward process will be executed per reconstruction. \par 

The model used in this work consists of several components, which we highlight below:

\textbf{Shower Profile} : The longitudinal profile of the air shower, describing the evolution of the air shower at each atmospheric depth $X$ (in units of grams per square centimetre), is parameterised by the Gaisser-Hillas function in the LR formalism, where the width and asymmetry parameters are described by $L$ and $R$ respectively~\cite{Andringa:2011zz}: 
\begin{equation}
    N(X) = N_\mathrm{max} \exp\left(-\dfrac{X - X_\mathrm{max}}{RL}\right) \: \left(1 + \dfrac{R}{L} (X - X_\mathrm{max})\right)^{R^{-2}}.
    \label{eq:gaisser_hillas}
\end{equation}
Here $N_\mathrm{max}$ is the number of electrons and positrons at $X_\mathrm{max}$ which is proportional to the electromagnetic energy of the cosmic ray. The implications of L and R on the mass separation between primary particles have been thoroughly studied in the context of the Pierre Auger Observatory~\cite{Andringa:2011zz} and SKAO~\cite{Corstanje:2023uyg,Buitink:2023reh}. Furthermore, their correlations are more uniform as compared to other standard parameterisations (such as with $X_0$ and $\lambda$). As such, it is a suitable choice for our model, and we can interpret the reconstructed shower parameters in a physically motivated manner. In order to capture fluctuations that cannot be expressed by the Gaisser-Hillas function, we also incorporate fluctuations based on the Ornstein-Uhlenbeck process at each atmospheric bin~\cite{Uhlenbeck:1930zz}. \par 

\begin{wrapfigure}[22]{L}{0.5\textwidth}
    \centering
    \includegraphics[width=0.5\textwidth]{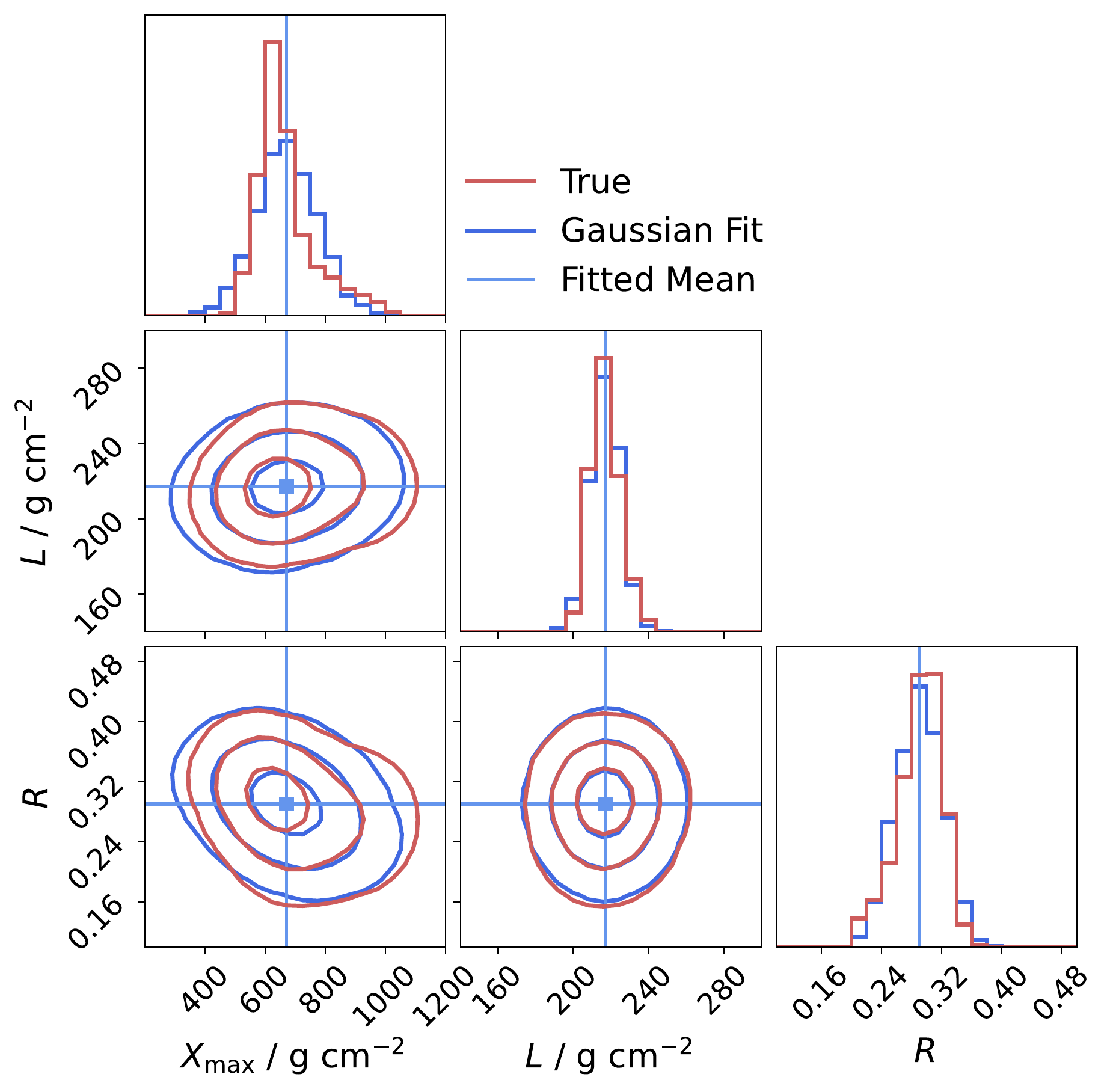}
    \caption{Prior distribution for $X_\mathrm{max}$, $L$, and $R$ used in this model. The true distribution obtained from \texttt{CORSIKA} simulations are also overlaid. The contours are plotted according to the 1, 2, and 3$\sigma$ confidence level for a 2-D Gaussian distribution. Figure generated using \texttt{corner.py} \cite{corner}.}
    \label{fig:shower_priors}
\end{wrapfigure}

In our model, we place prior distributions on each shower parameter that encodes our physical understanding of each parameter. Previous studies have shown that $X_\mathrm{max}$, $L$ and $R$ exhibit strong correlations between each other. As such, we describe the shower parameters by a multivariate Gaussian with mean and covariance matrix that encodes the correlations between the parameters. The mean and covariance are estimated from shower parameters obtained from 10,000 pre-generated \texttt{CORSIKA} simulations~\cite{Heck:1998vt} with the hadronic interaction model QGSJETII-04~\cite{Ostapchenko2013} specific to atmospheric conditions in the context of LOFAR~\cite{Corstanje:2021kik}. \Cref{fig:shower_priors} shows a corner plot of the prior distribution of the shower parameters overlaid on the true distribution obtained from \texttt{CORSIKA} events. We also apply physical limits of $X_\mathrm{max} \in [400, 1100]$\,g/cm$^2$, $L \in [200, 240]$\,g/cm$^2$, and $R \in [0.2, 0.4]$. As $N_\mathrm{max}$ is only weakly correlated with the other shower parameters, we set a normal prior $\log_{10}(N_\mathrm{max}) \sim \mathrm{Normal}(8, 0.5)$ independent of other parameters. \par 

\begin{figure}
    \centering
    \includegraphics[width=0.65\textwidth]{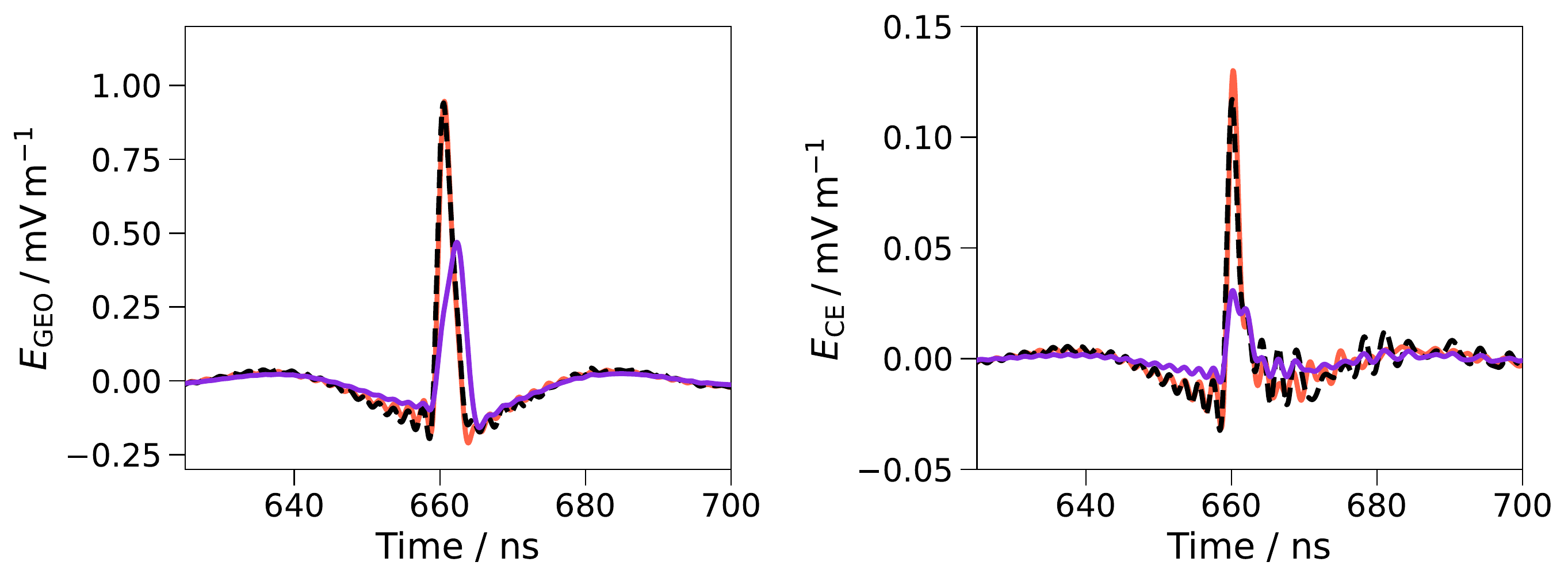}
    \includegraphics[width=0.3\textwidth]{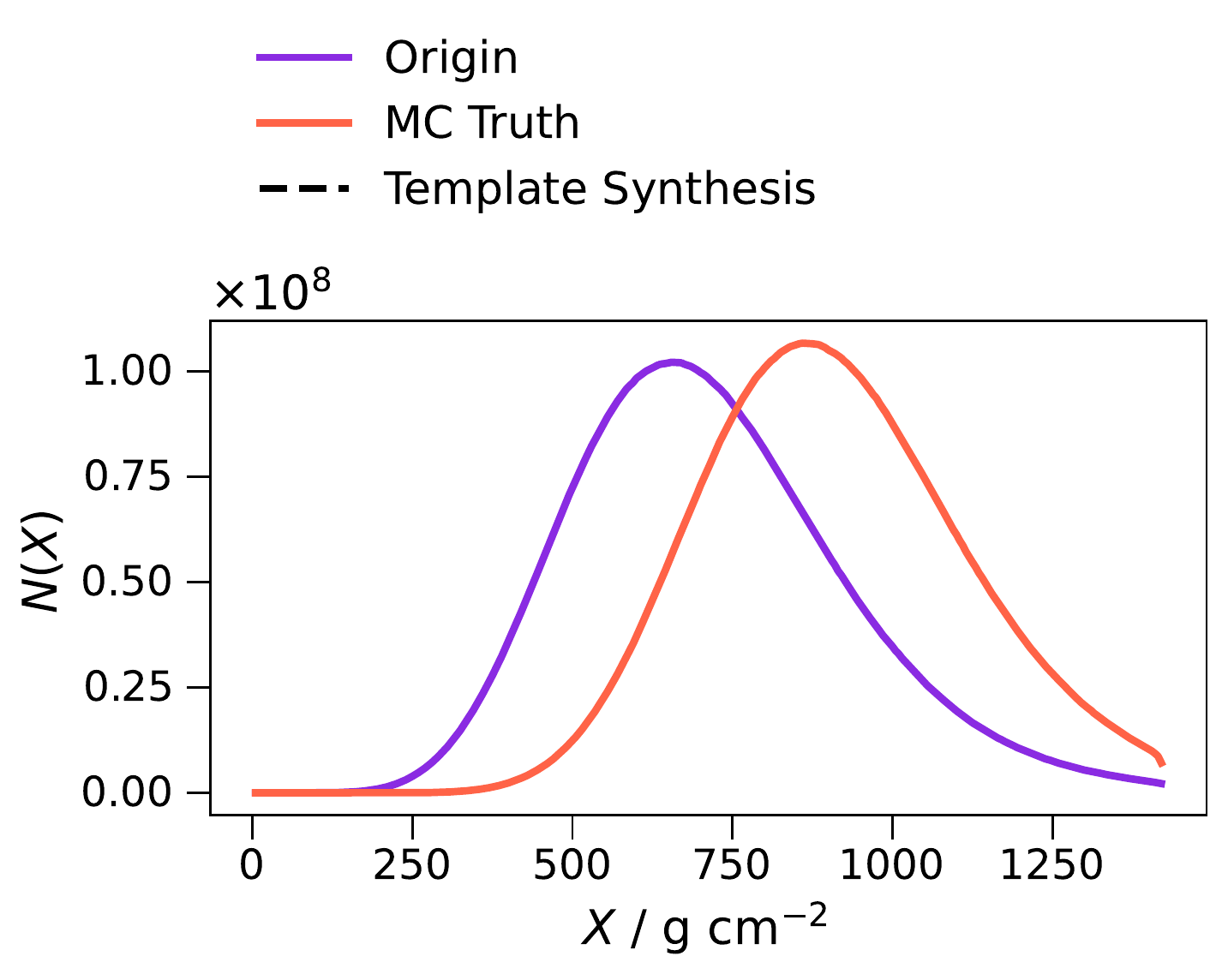}
    \caption{Left: An example of a synthesised trace using template synthesis (black, dashed) within a frequency bandwidth of $[30, 500]$\,MHz at an antenna position $d_\mathrm{core}=\SI{52}{\meter}$ from the shower core. We apply a template with an origin shower (overlaid in purple) with a zenith and azimuth angle of $\theta=42^\circ$, $\phi=117^\circ$ respectively. We also show the electric field traces generated from a \texttt{CoREAS} simulation with the same configuration (red) to verify its accuracy. Right: The longitudinal profile of the origin (purple) and synthesised shower (red), where we use the same profile from \texttt{CoREAS} for the synthesised shower. }
    \label{fig:template_synthesis}
% \end{wrapfigure}
\end{figure}

\textbf{Radio Emission Model} : In order to quickly but accurately model the radio emission generated from a given shower profile, we use \texttt{SMIET}~\cite{Desmet:2025ufy}, which allows us to synthesise the radio pulses from any longitudinal profile using pre-computed templates of the electric field signal at each shower geometry for a fixed antenna layout within a matter of seconds. The resulting electric fields are described by the geomagnetic and charge-excess emission that encapsulates the physical mechanisms of the radio emission. The template is based on an ``origin shower'', that is, a generated shower based on \texttt{CoREAS} simulations which simulate the electric field signal at fixed antenna positions from individual atmospheric slices. Using scaling relations previously and independently pre-computed from \texttt{CoREAS} simulations, we can re-scale the origin shower to generate the desired template. Applying this template to any given shower profile, we can synthesise the electric field pulse at each antenna position. This approach matches the accuracy of \texttt{CoREAS} simulations within 6\% for $\Delta X_\mathrm{max} \leq \SI{100}{\gram\per\centi\meter\squared}$ from the origin shower. \Cref{fig:template_synthesis} shows an example of a synthesised trace with \texttt{SMIET} compared to the true electric field signal as obtained from a \texttt{CoREAS} simulation with the same configuration. \par  

\texttt{SMIET} an ideal forward model for our reconstruction framework, and its development was motivated by the aim to apply it in an IFT reconstruction. Firstly, we can describe the radio emission from any profile without the need for extensive simulations for every possible shower geometry. Furthermore, each synthesis process can be performed on the order of seconds, vastly increasing the speed of the forward model compared to running \texttt{CoREAS} simulations. Lastly, its semi-analytical nature makes the model fully differentiable, which is a necessary requirement for the IFT framework. While template synthesis can be applied to other shower geometries through the interpolation of the phase spectrum, we opt to currently use only a fixed geometry for our model; we will extend our model to allow variations of the arrival direction in future works. \par 

\textbf{Antenna \& Noise Model} : The antenna response describes how the electric field signal is detected at each antenna, which depends on the antenna type and orientation. The electric fields can be transformed into the voltage traces observed at each antenna through the following response function $H_{ij}$:
\begin{equation}
    \begin{bmatrix}
        V_X \\ V_Y 
    \end{bmatrix}
    = 
    \begin{bmatrix}
        H_{X, \theta} & H_{X, \phi} \\
        H_{Y, \theta} & H_{Y, \phi} \\
    \end{bmatrix}
    \cdot 
    \begin{bmatrix}
        E_\theta \\ E_\phi 
    \end{bmatrix}
\label{eq:antenna_response}
\end{equation}
In our model, we use the response function modelled for the SKA-Low Log-periodic Antennas (SKALA4)~\cite{Bolli2020}, implemented through the \texttt{NuRadioReco} software framework~\cite{Glaser:2019rxw}. We apply a frequency bandwidth of $[50, 200]$\,MHz with a sampling rate of $\SI{500}{\mega\hertz}$, which adequately captures the detector response of the SKA-low antennas. As \texttt{SMIET} synthesises pulses at fixed antenna positions, we only work with fixed star-shaped antenna layouts for now and plan to utilise a realistic antenna layout from SKA-Low through pulse interpolation~\cite{Corstanje:2023vqp} in future works. \par 

The noise within the signal has a variety of sources, of which the most prominent are the Galactic radio emission and instrumental noise. While it is possible to model these noise sources separately, we characterise the combined broadband noise through a normal distribution with an RMS voltage of $\sigma_N^V = \SI{7e-5}{\volt}$ such that all events used for our reconstruction have a signal-to-noise ratio (SNR), defined by the ratio of the maximum amplitude of the trace to the noise RMS, to be greater than one. 

\section{Results}

To verify our model, we compare our results with simulated events used in~\cite{Corstanje:2021kik} which have been generated using \texttt{CoREAS} with environmental conditions for the LOFAR site. From this dataset, we take seven event sets with varying energies and event geometries (zenith and azimuth angles) with protons as their primary particle. Each set consists of approximately 20 simulations that we use for our reconstruction. For each set, we then generate a single origin shower with $X_\mathrm{max}^\mathrm{origin} \geq \SI{640}{\gram\per\centi\meter\squared}$. Each origin shower is then used within our model through template synthesis. The event sets used in this work are listed in \Cref{tab:event_config}.  \par 

% \begin{wraptable}{r}{0.8\textwidth}
\begin{table}
    \centering
    \begin{tabular}{|c|c|c|c|c|c|c|}
    \hline
        ID & $\theta$ / deg & $\phi$ / deg & $E$ / $10^{17}$ eV & $X_\mathrm{max}^\mathrm{origin}$ / $\si{\gram\per\centi\meter\squared}$ & $N_\mathrm{sims}$ \\ \hline
        65490891 & 28.5 & 85.3 & 2.04 & 642 & 19 \\
        62804598 & 42.8 & 117.1 & 1.53 & 655 & 21 \\
        82321543 & 39.7 & 135.9 & 3.73 & 678 & 21 \\
        81409140 & 26.1 & 223.1 & 7.00 & 714 & 18 \\
        70360905 & 21.0 & 39.7 & 1.83 & 723 & 21 \\
        64021966 & 31.3 & 319.1 & 2.15 & 734 & 20 \\
        72094854 & 30.4 & 122.9 & 7.98 & 803 & 22 \\ \hline
    \end{tabular}
    \caption{Event sets (zenith, azimuth, energy, and $X_\mathrm{max}$ for the origin shower) used in this work. The number of simulations contained per event is also listed.}
    \label{tab:event_config}
% \end{wraptable}
\end{table}

The simulated events are convolved with the antenna response matrix to generate the detected voltage pulse, to which we then add a random noise signal at each time bin generated from our noise model to synthetically generate data. In this way, we generate a mock signal that incorporates both broadband noise and the voltage signal. Using this data, generated for each event, we apply our model to infer the shower parameters and benchmark against the true values for each simulation. \par

\begin{figure}
    \centering
    \includegraphics[width=0.42\textwidth]{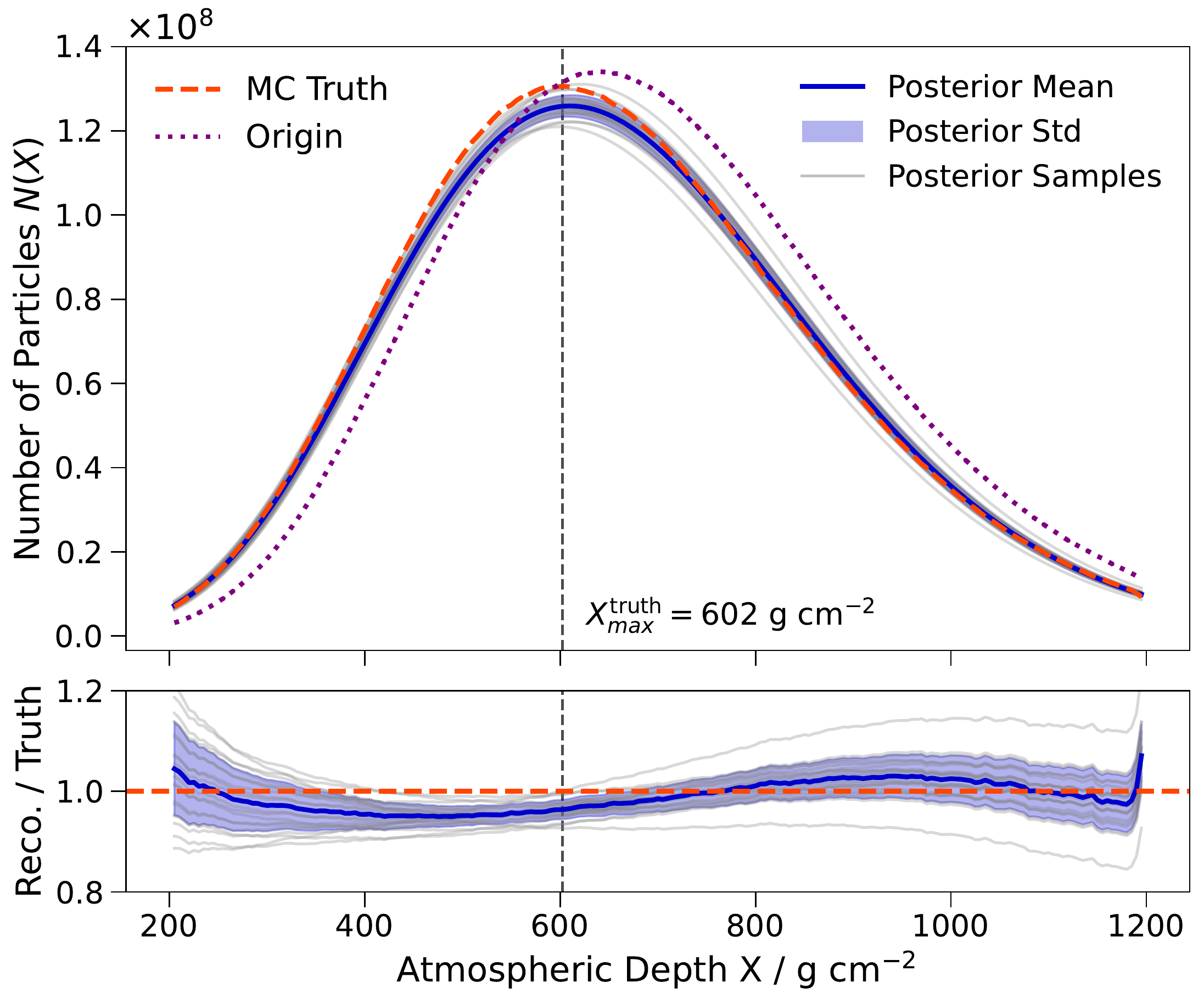}
    \includegraphics[width=0.46\textwidth]{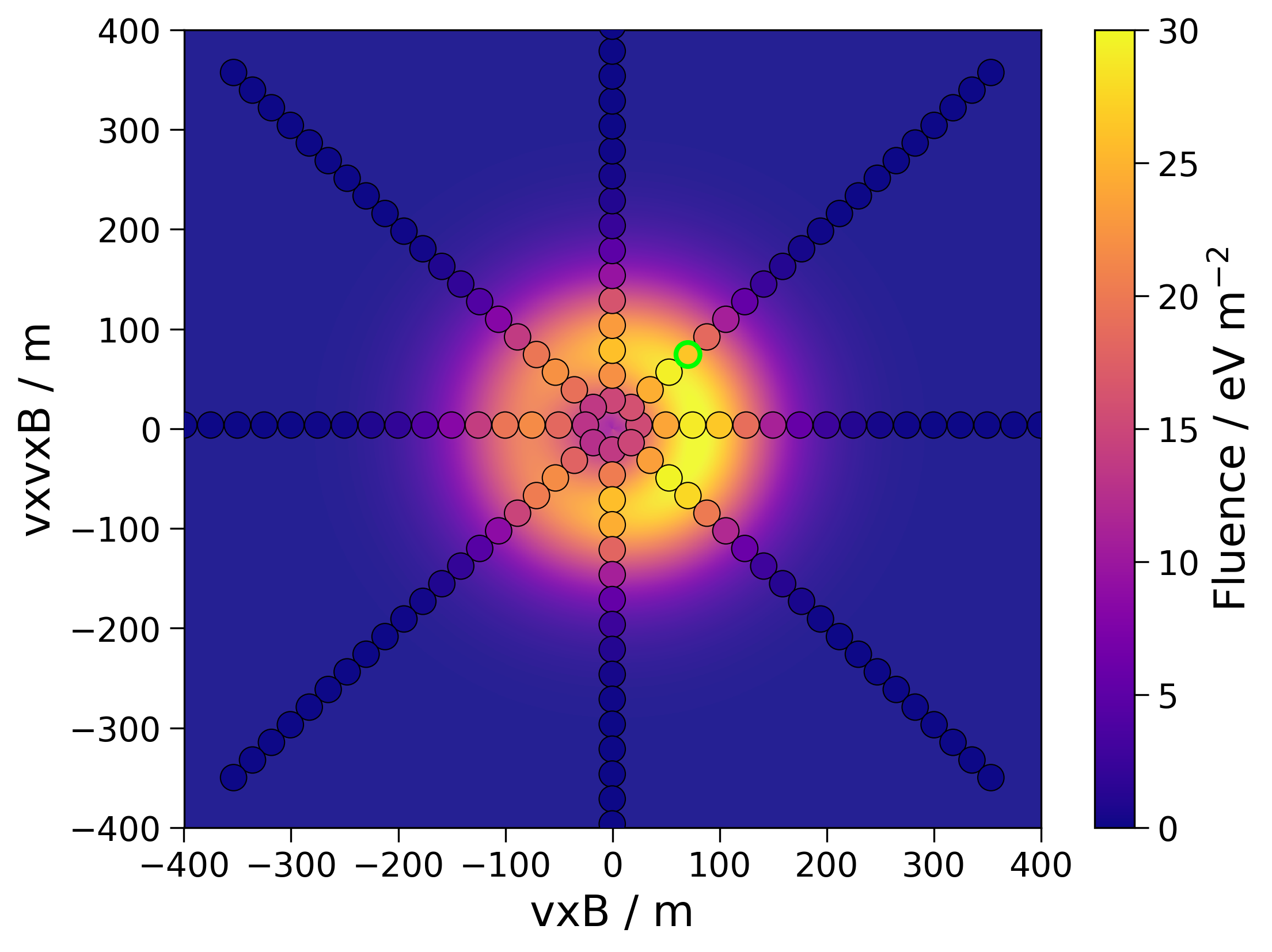} \\
    
    \includegraphics[width=0.45\textwidth]{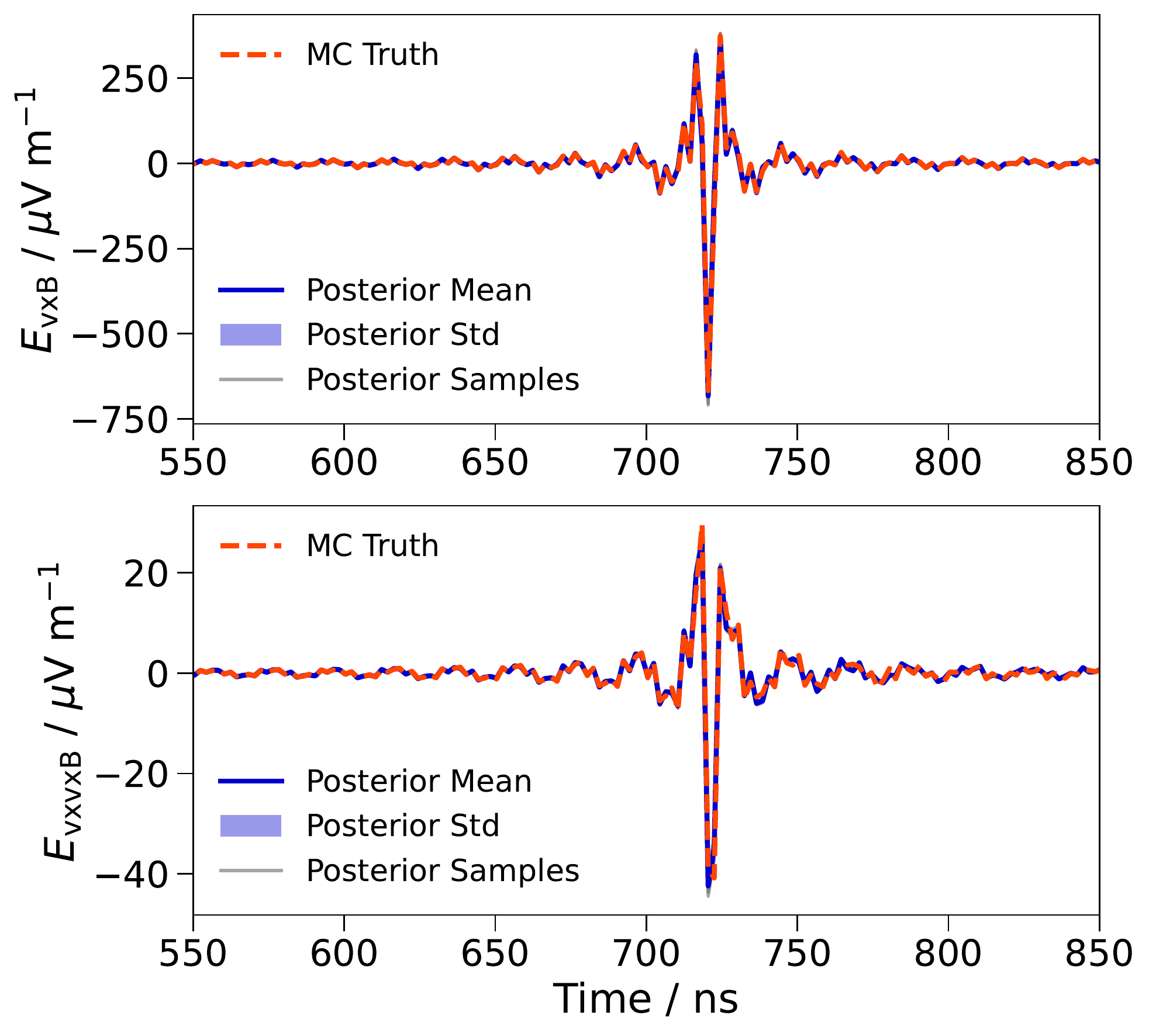}
    \includegraphics[width=0.45\textwidth]{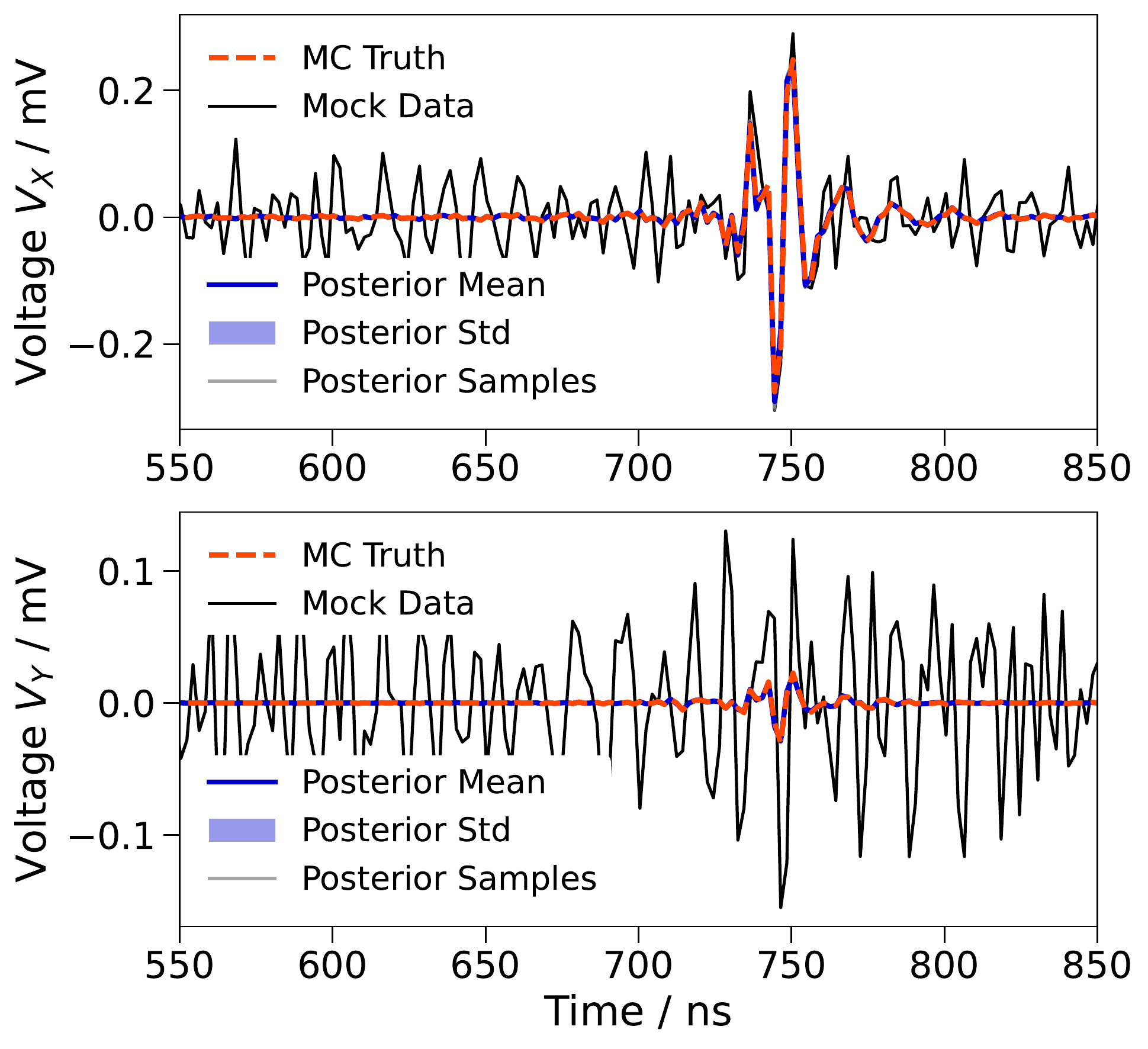}
    \caption{An example of an event reconstruction from this model for the event configuration 65490891 with $X_\mathrm{max}^\mathrm{truth} = \SI{602}{\gram\per\centi\meter\squared}$. Top Left: The posterior mean and standard deviation (blue) of the longitudinal profile, compared to the true profile from the \texttt{CoREAS} simulation (red, dashed). Each posterior sample used for inference (gray) and the profile from the origin shower used in the template (purple, dotted) is also shown. The ratio between the true and reconstructed profiles are shown in the subplot.
    Top Right: The posterior mean of the energy fluence at each antenna (depicted by circles) in the shower plane. The interpolated true energy fluence is overlaid to show the qualitative agreement with the reconstructed results.
    Bottom Left: The reconstructed electric field traces in the shower plane at an antenna $d_\mathrm{core} = \SI{106}{\meter}$ away from the shower core (indicated by the green circle on the fluence map). 
    Bottom Right: The reconstructed voltage traces at the same antenna position in the X/Y polarisation of the SKALA antenna. The mock data used for reconstruction is also shown. }
    \label{fig:event_reco}
\end{figure}

\Cref{fig:event_reco} shows an example of an event reconstruction, where we show the reconstructed longitudinal profile, energy fluence, electric field, and voltage traces for an event with $X_\mathrm{max}^\mathrm{truth} = \SI{602}{\gram\per\centi\meter\squared}$, taking event configuration 65490891 as listed in \Cref{tab:event_config}. For the electric field and voltage traces, the pulses at an antenna position with $d_\mathrm{core} = \SI{106}{\meter}$ away from the shower core were taken (highlighted in green in the figure). We show that the longitudinal profile can be reconstructed within one standard deviation, except at atmospheric depths from approximately $\SI{400}{\gram\per\centi\meter\squared}$ to $\SI{700}{\gram\per\centi\meter\squared}$, where the profile is underestimated. We also observe that the errors increase near the edges of the atmospheric grid, which is expected as the number of particles decreases significantly in such regions. We highlight that, through our model, we can also reconstruct both electric field and voltage signals. In particular, we can recover the amplitude and shape of the pulse for most antenna positions. For antenna positions along the vxB axis, where we approximate the emission on the vxB plane in our model by averaging the emission from nearby arms, the reconstructed fluence is slightly underestimated.

\begin{figure}
    \centering
    \includegraphics[width=0.8\textwidth]{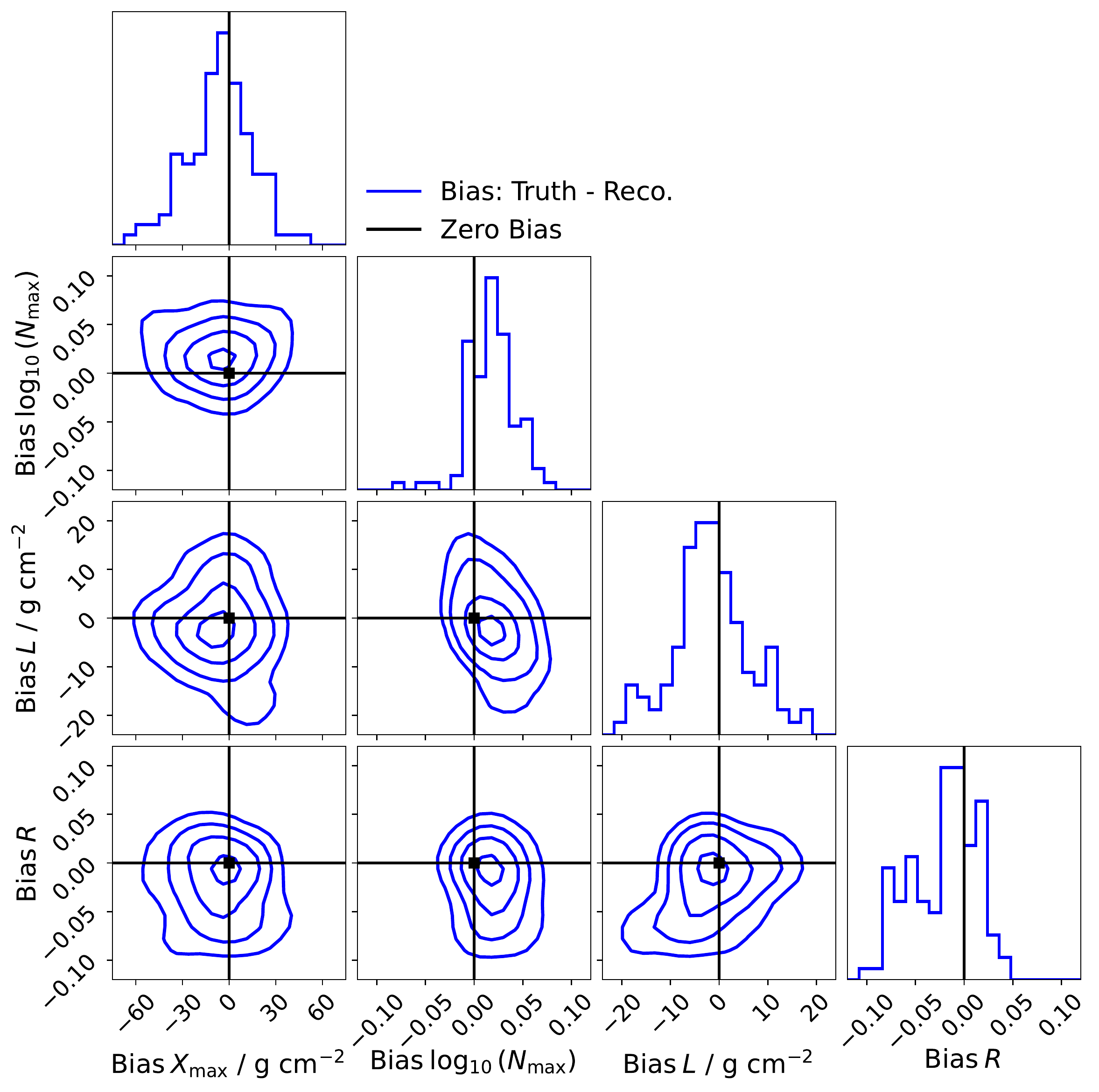}
    \caption{Distribution of the bias of the shower parameters $X_\mathrm{max}$, $\log_{10} (N_\mathrm{max})$, $L$ and $R$ for all events considered in this work. The contours show the 0.5, 1, 1.5, and 2$\sigma$ values following a 2-D Gaussian distribution. The point where the bias is zero is indicated by the black lines and dots. Figure generated using \texttt{corner.py}~\cite{corner}.}
    \label{fig:bias_shower_params}
\end{figure}

To verify our model for all events, \Cref{fig:bias_shower_params} shows the distribution of the bias (difference between Monte-Carlo true parameter and reconstructed value) for each reconstructed shower parameter ($X_\mathrm{max}$, $\log_{10} N_\mathrm{max}$, $L$, and $R$). We also display the correlations between the biases of each parameter. We show that on average most reconstructed shower parameters are unbiased, however with a somewhat large variation. In particular, the precision of $X_\mathrm{max}$ amounts to $\SI{25}{\gram\per\centi\meter\squared}$, which is still slightly larger than the reported precision from LOFAR \cite{Buitink:2014} amounting to $\SI{19}{\gram\per\centi\meter\squared}$. Furthermore, we observe that our model currently underestimates $N_\mathrm{max}$, which can also be observed from the underestimated profile reconstruction around $X_\mathrm{max}$. The origin of this bias is not yet known and is under investigation. We also observe that some events overestimate the asymmetry parameter $R$, which is loosely correlated to the reconstruction efficiency of the shower width $L$. This correlation is expected, as both $L$ and $R$ are strongly dependent on each other~\cite{Andringa:2011zz}. \par 

\section{Conclusion}

In this work we showcase the development of a novel method to reconstruct the longitudinal profile of air showers using radio measurements. The reconstruction is performed via Information Field Theory, a Bayesian framework applicable for signal reconstruction. We parameterise the shower profile using the Gaisser-Hillas function in the LR formalism. The radio emission observed at each antenna is modelled using template synthesis, where electric field pulses are synthesised for any profile with a given simulated shower. We generate voltage traces by convolving with the antenna response from the SKALA4 antennas and apply Gaussian noise to each time bin. We verified our reconstruction framework with seven event configurations with an ideal star-shaped antenna layout, showcasing that, while we currently underestimate the particle number, we are able to recover the longitudinal profile within the 1$\sigma$ uncertainty band. The depth of shower maximum $X_\mathrm{max}$ is reconstructed without a bias and with a resolution of currently $\SI{25}{\gram\per\centi\meter\squared}$.  Furthermore, we showcase that we also reconstruct the electric field and voltage traces, recovering both their shape and amplitudes.  \par 

In this work, we tested our model with approximately 100 simulations. To properly quantify the performance of our model, we plan to verify our reconstruction framework with all $\sim$20\,000 events within the same dataset. We also plan to extend this model by using pulse interpolation to apply our framework on realistic antenna layouts for LOFAR and SKA-Low. Furthermore, we hope to include the interpolation of the phase spectrum within template synthesis to perform arrival direction variation along with the shower profile, potentially lifting the restriction of single event geometry reconstructions.

\let\oldbibliography\thebibliography
\renewcommand{\thebibliography}[1]{%
  \oldbibliography{#1}%
  \setlength{\itemsep}{1pt}%
}

{\footnotesize
\bibliographystyle{JHEP}
\bibliography{publications}
}
\newpage
\newcommand{\affilASTRON}{Netherlands Institute for Radio Astronomy (ASTRON), Dwingeloo, The Netherlands}
\newcommand{\affilCanTho}{Physics Education Department, School of Education, Can Tho University, Campus~II, 3/2 Street, Ninh Kieu District, Can Tho City, Viet Nam}
\newcommand{\affilCurtin}{International Centre for Radio Astronomy Research, Curtin University, Bentley, 6102, WA, Australia}
\newcommand{\affilDESY}{Deutsches Elektronen-Synchrotron DESY, Platanenallee~6, 15738 Zeuthen, Germany}
\newcommand{\affilErlangen}{Erlangen Centre for Astroparticle Physics, Friedrich-Alexander-Universit\"at Erlangen-N\"urnberg, 91058 Erlangen, Germany}
\newcommand{\affilGorlitz}{Deutsches Zentrum f\"ur Astrophysik, Postplatz~1, 02826 Görlitz, Germany}
\newcommand{\affilGroningen}{Kapteyn Astronomical Institute, University of Groningen, P.O.~Box 72, 9700 AB Groningen, Netherlands}
\newcommand{\affilHefei}{School of Astronomy and Space Science, University of Science and Technology of China, Hefei 230026, China}
\newcommand{\affilKanpur}{Department of Physics, Indian Institute of Technology Kanpur, Kanpur, UP-208016, India}
\newcommand{\affilKeyNanjing}{Key Laboratory of Modern Astronomy and Astrophysics, Nanjing University, Ministry of Education, Nanjing 210023, China}
\newcommand{\affilKIT}{Institut f\"ur Astroteilchenphysik, Karlsruhe Institute of Technology (KIT), P.O.~Box 3640, 76021 Karlsruhe, Germany}
\newcommand{\affilKhalifa}{Department of Physics, Khalifa University, P.O.~Box 127788, Abu Dhabi, United Arab Emirates}
\newcommand{\affilManchester}{Jodrell Bank Centre for Astrophysics, Department of Physics and Astronomy, University of Manchester, Manchester M13 9PL, UK}
\newcommand{\affilMaxPlanck}{Max-Planck Institut f\"ur Astrophysik, Karl-Schwarzschild-Str.~1, 85748 Garching, Germany}
\newcommand{\affilMunich}{Ludwig-Maximilians-Universit\"at M\"unchen (LMU), Geschwister-Scholl-Platz~1, 80539 M\"unchen, Germany}
\newcommand{\affilNanjing}{School of Astronomy and Space Science, Nanjing University, Nanjing 210023, China}
\newcommand{\affilNijmegen}{Department of Astrophysics/IMAPP, Radboud University Nijmegen, P.O.~Box 9010, 6500 GL Nijmegen, The Netherlands}
\newcommand{\affilNikhef}{Nikhef, Science Park Amsterdam, 1098 XG Amsterdam, The Netherlands}
\newcommand{\affilPurpleMt}{Key Laboratory of Dark Matter and Space Astronomy, Purple Mountain Observatory, Chinese Academy of Sciences, No.~10 Yuanhua Road, Nanjing, China}
\newcommand{\affilULB}{Universit\'e Libre de Bruxelles, Science Faculty CP230, B-1050 Brussels, Belgium}
\newcommand{\affilVUB}{Vrije Universiteit Brussel, Astrophysical Institute, Pleinlaan~2, 1050 Brussels, Belgium}
\newcommand{\affilXidian}{School of Electronic Engineering, Xidian University, No.2 South Taibai Road, Xi'an, China}
\newcommand{\affilSKA}{SKA Observatory, Jodrell Bank, Lower Withington, Macclesfield, SK11 9FT, UK}

\section*{Affiliations}
\scriptsize
\noindent
$^a$ {\affilKIT} \\
$^b$ {\affilErlangen} \\
$^c$ {\affilManchester} \\
$^d$ {\affilVUB} \\
$^e$ {\affilNijmegen} \\
$^f$ {\affilCurtin} \\
$^g$ {\affilMaxPlanck} \\
$^h$ {\affilMunich} \\
$^i$ {\affilGorlitz} \\
$^j$ {\affilGroningen} \\
$^k$ {\affilASTRON} \\
$^l$ {\affilPurpleMt} \\
$^m$ {\affilNikhef} \\
$^n$ {\affilDESY} \\
$^o$ {\affilKanpur} \\
$^p$ {\affilKhalifa} \\
$^q$ {\affilCanTho} \\
$^r$ {\affilSKA} \\
$^s$ {\affilNanjing} \\
$^t$ {\affilKeyNanjing} \\
$^u$ {\affilXidian} \\
$^v$ {\affilHefei} \\

\section*{Acknowledgements} \noindent
SBo, AN and KT acknowledge funding through the Verbundforschung of the German Federal Ministry of Research, Technology and Space (BMFTR). PL, KW and MJ are supported by the Deutsche Forschungsgemeinschaft (DFG, German Research Foundation) – Projektnummer 531213488. MD is supported by the Flemish Foundation for Scientific Research (FWO-AL991). ST acknowledges funding from the Khalifa University RIG-S-2023-070 grant. SB acknowledges funding from the Medium-Scale Infrastructure program of the Flemish Foundation for Scientific Research (FWO). KM acknowledges funding from the Netherlands Research School for Astronomy (NOVA) Phase 6 Instrumentation Call.  The authors gratefully acknowledge the computing time provided on the high-performance computer HoreKa by the National High-Performance Computing Center at KIT (NHR@KIT). This center is jointly supported by the Federal Ministry of Education and Research and the Ministry of Science, Research and the Arts of Baden-W\"{u}rttemberg, as part of the National High-Performance Computing (NHR) joint funding program. HoreKa is partly funded by the German Research Foundation.

\end{document}